\documentclass[floatfix, preprint, onecolumn, prd,showpacs, showkeys, preprintnumbers, nofootinbib, superscriptaddress]{revtex4-1}
\usepackage{times}
\usepackage{amssymb,amsbsy,amsmath,amsfonts}
\usepackage{graphicx}
\usepackage{float}
\usepackage{color}
\usepackage{morefloats}
\usepackage{rotating}
\usepackage{srcltx}
\usepackage{slashed}
\usepackage{multirow}
\usepackage{verbatim}
\usepackage{hyperref}
\usepackage{tabularx}

\begin{document}

\title{Novel nonperturbative approach for radiative $\bar{B}^0(\bar{B}^0_s)\rightarrow J/\psi \gamma$ decays}

\author{Li-Sheng Geng} \email{lisheng.geng@buaa.edu.cn}
\affiliation{School of Physics and Nuclear Energy Engineering and
International Research Center for Nuclei and Particles in the
Cosmos, Beihang University, Beijing 100191, China}
\affiliation{Beijing Key Laboratory of Advanced Nuclear Materials and Physics, Beihang University, Beijing 100191, China}

\author{Eulogio Oset}
\affiliation{Departamento de F\'{\i}sica Te\'orica and IFIC, Centro Mixto Universidad de Valencia-CSIC Institutos de Investigaci\'on de Paterna, Aptdo. 22085, 46071 Valencia, Spain
}

\date{\today}

\begin{abstract}
Radiative $\bar{B}^0(\bar{B}^0_s)\rightarrow J/\psi \gamma$ decays provide an interesting case to test our understanding of (non)perturbative QCD and eventually to probe physics beyond the
standard model.  Recently, the LHCb Collaboration has reported an upper bound, updating the results of the BABAR Collaboration. Previous theoretical predictions
based on QCD factorization or perturbative QCD have shown large variations due to different treatment of 
nonfactorizable contributions and meson-photon transitions. In this paper, we report on a novel approach to estimate the decay rates, which is based on
a recently proposed model for $B$ decays and the vector meson dominance hypothesis, widely tested in the relevant energy regions. The predicted branching ratios 
are $\mathrm{Br}[\bar{B}^0\rightarrow J/\psi\gamma]=\left(3.50\pm0.34^{+1.12}_{-0.63}\right)\times10^{-8}$ and  
$\mathrm{Br}[\bar{B}^0_s\rightarrow J/\psi\gamma]=\left(7.20\pm0.68^{+2.31}_{-1.30}\right)\times10^{-7}$. The first uncertainty is systematic and the second is statistical, originating from
the experimental $\bar{B}^0_s\rightarrow J/\psi \phi$ branching ratio.  

\end{abstract}

\maketitle
\section{Introduction}
The study of weak $B$ decays is turning into an unexpected very valuable source of information on hadron dynamics \cite{sheldontalk,review}. The $B$ and $B_s$ decays into $J/\psi$ and a pair of pions brought surprises showing that  in the $B^0_s$ decay the two pions produced a big signal of the $f_0(980)$ resonance and no trace of the $f_0(500)$ \cite{Aaij:2011fx,LHCb:2012ae,Aaij:2014emv,Li:2011pg,Aaltonen:2011nk,Abazov:2011hv}, while in the $B^0$ decay the two pions contributed strongly in the $f_0(500)$ region and made only a small contribution in the  $f_0(980)$ region \cite{Aaij:2013zpt,Aaij:2014siy}. A study of these processes taking into account explicitly the final state interaction of the pions, together with its coupled channels, gave a good interpretation of these features \cite{liang}, providing support for the picture of the chiral unitary approach, where these two resonances emerge as a consequence of the interaction of pseudoscalar mesons in coupled channels \cite{npa,ramonet,kaiser,markushin,juanito,rios,sigma}. Similarly, the study of $B$ and $B_s$ decays into $J/\psi$ and a vector meson \cite{bayarliang} gave support to the picture in which the vector mesons are basically made of $q \bar q$ \cite{rios,sigma}.

  For the $B (B_s) \to J/\psi \gamma$ decays, only upper bounds for their branching ratios of the order of $10^{-6}$ are available so far \cite{Aubert:2004xd,lhcb}.
Theoretical studies of these decays are available and they use the naive factorization, or QCD factorization \cite{luwang}, perturbative QCD with the $K_T$ factorization \cite{lilu}, or other kinds of factorization approximations \cite{anas}. In the present paper we address the problem in a different way, establishing a link to the $B (B_s) \to J/\psi V$ decays, with V a vector meson, which were studied in \cite{bayarliang}. The link to the $B (B_s) \to J/\psi \gamma$ is then established by converting the vector meson V into a photon, using for it the vector meson dominance (VMD) hypothesis \cite{sakurai}, which is most practically implemented using the local hidden gauge approach \cite{hidden1,hidden2,hidden4}. The intricate form factors stemming from the weak decay and QCD matrix elements are taken into account by using the experimental value of the $B^0_s \to J/\psi \phi$ decay width. This new way of addressing the problem provides rates which should be rather accurate, and they come at a moment where the rates obtained from the other theoretical approaches differ sometimes by two orders of magnitude. Also  significant is the fact that, while the rates obtained are below the present upper bounds, the branching ratio for  $B^0_s \to J/\psi \gamma$ is only one order of magnitude smaller than the present experimental bound. This should serve as a motivation to push the experimental limits to get absolute values for this rate that could shed some light on the theoretical methods to address the problem. More problematic is the $B^0 \to J/\psi \gamma$ decay, where we predict a rate about two orders of magnitude smaller than  the experimental bound, but should this rate be measured it would also help us understand better the relevant physics behind these processes.

 The paper is organized as follows. In the next section we present the formalism. Sec. III considers further theoretical uncertainties and compares
 the final results with those of other theoretical approaches. We finish with some conclusions in Sec. IV.
 
\section{Formalism}
The idea that we follow is to link the $\bar{B}^0\rightarrow J/\psi\gamma$ and $\bar{B}^0_s\rightarrow J/\psi\gamma$ reactions to 
$\bar{B}^0\rightarrow J/\psi\rho$,  $J/\psi\omega$, and $\bar{B}^0_s\rightarrow J/\psi\phi$ by converting
the neutral vector mesons into a photon.  For this purpose we use the Sakurai VMD hypothesis~\cite{sakurai}, which is most practically implemented using the local hidden gauge
approach~\cite{hidden1, hidden2,hidden4}. The $\bar{B}^0(\bar{B}_s^0)\rightarrow J/\psi V$ production is addressed following the work of \cite{bayarliang}, which we describe briefly below.

\begin{figure}
  \centering
  \includegraphics[width=0.4\textwidth]{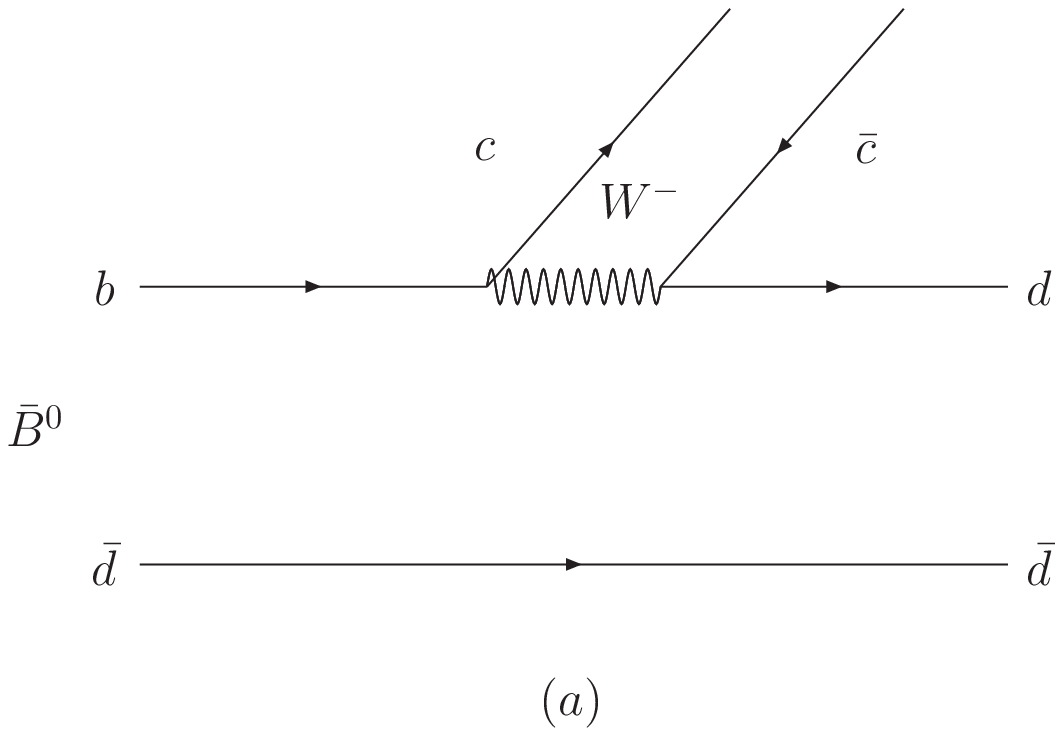}\;\;\;
 \includegraphics[width=0.4\textwidth]{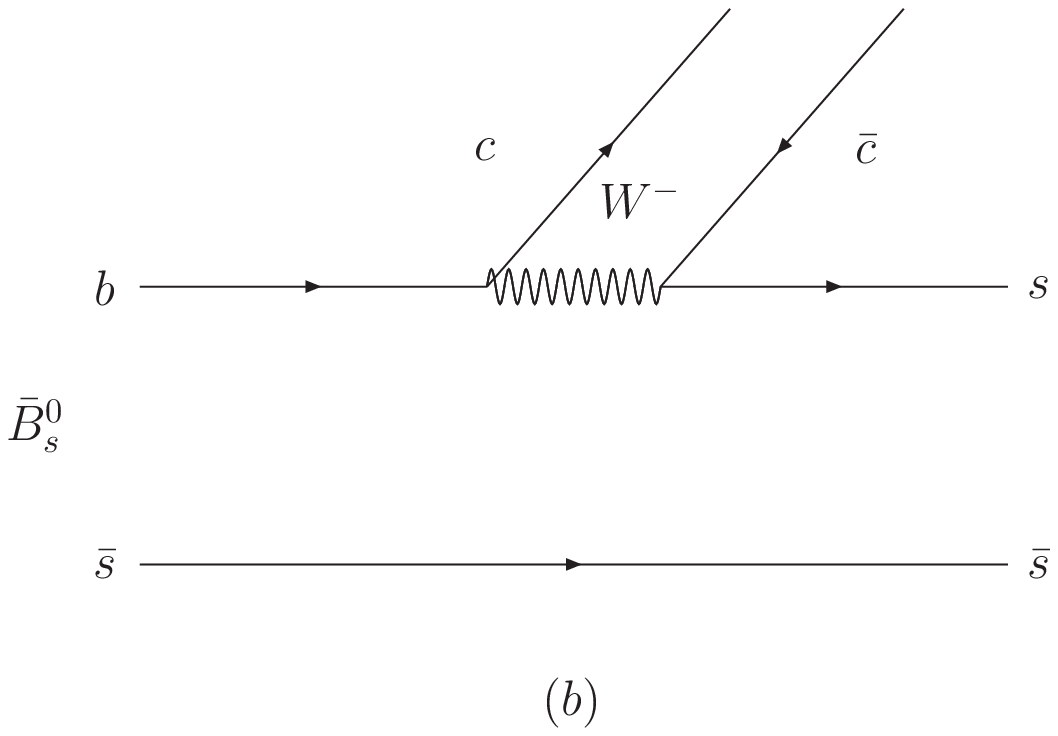}
  \caption{Diagrams for (a) $\bar{B}^0$ and (b) $\bar{B}^0_s$  decays into $c\bar{c}$, making $J/\psi$, and a pair of light quarks.}\label{bdecay}
\end{figure}

The $\bar{B}^0$ and $\bar{B}^0_s$ decay mechanisms at the quark level are given in Fig.~\ref{bdecay}~\cite{stone,liang}. The first thing to note is that in diagram (a) for the $\bar{B}^0$ decay one has the quark 
transition $cd$, which requires the Cabibbo-Kobayashi-Maskawa (CKM) matrix element, $V_{cd}=\sin\theta_c$, and hence is Cabibbo suppressed. On the other hand, in the decay of
$\bar{B}_s^0$, diagram (b), one has the $cs$ transition that goes with the CKM matrix element $V_{cs}=\cos\theta_c$, and hence the process is Cabibbo favored. In both
decays we create a $c\bar{c}$ pair that will make the $J/\psi$ meson and an extra pair of quarks, $d\bar{d}$ in the $\bar{B}^0$ decay and $s\bar{s}$ in the $\bar{B}^0_s$ decay. In \cite{liang}
this extra pair of quarks is hadronized, including a further $q\bar{q}$ pair with the quantum numbers of the vacuum, in order to have two mesons in the final state, in addition to 
the $J/\psi$. But here we are interested in the production of $\rho$, $\omega$, $\phi$ in addition to the $J/\psi$. Then it is most opportune to mention that the studies conducted to determine
the nature of mesons in terms of quarks conclude that the low-lying  scalar mesons come from the interaction of pseudoscalars,  but the low-lying vector mesons are basically $q\bar{q}$ states. This has been
thoroughly tested by using large $N_c$ arguments in \cite{rios} or applying an extension of the compositeness Weinberg sum rule~\cite{weinberg,hanhart} to states not so close to threshold~\cite{danijuan,hyodo, sekihara} and in particular in $\ell\ne 0$ partial waves~\cite{aceti,xiao}. Indeed, in \cite{aceti}, using experimental data and the generalized sum rule, one concludes that the amount of $\pi\pi$ in the
$\rho$ wave function is very small, of the order of 10\%. The same conclusion is obtained for the $K\pi$ component in the $K^*$ in \cite{xiao}. This means that the $q\bar{q}$ component  is the basic one in
the wave function of the vector mesons and we shall adhere to this picture.  In view of this, from Fig.~\ref{bdecay}, we can already describe the production of the $\rho$, $\omega$, and $\phi$ mesons in addition to the
$J/\psi$.  For this we recall that the wave functions of these mesons in terms of quarks are given by
\begin{eqnarray}
\rho^0&=&\frac{1}{\sqrt{2}}(u\bar{u}-d\bar{d}),\\
\omega&=&\frac{1}{\sqrt{2}}(u\bar{u}+d\bar{d}),\\
\phi&=&s\bar{s}.
\end{eqnarray}

Next, as done in \cite{liang,bayarliang}, we refrain from doing an elaborate evaluation of the matrix elements involved in the weak decay and factorize them in
terms of a factor $V'_p$, in view of which, the amplitudes for $J/\psi V$ production are given by
\begin{eqnarray}
t_{\bar{B}^0\rightarrow J/\psi\rho^0}&=&-\frac{1}{\sqrt{2}}V'_p V_{cd},\label{BJV1}\\
t_{\bar{B}^0\rightarrow J/\psi\omega}&=&\frac{1}{\sqrt{2}}V'_p V_{cd},\label{BJV2}\\
t_{\bar{B}^0_s\rightarrow J/\psi\phi}&=& V'_p V_{cs},\label{BJV3}
\end{eqnarray}
where we have explicitly spelled out the CKM matrix elements that distinguish one process from the other.

In \cite{bayarliang} it was shown that using Eqs.~(\ref{BJV1})--(\ref{BJV3}) and similar ones for $\bar{B}^0\rightarrow J/\psi \bar{K}^{*0}$ and $\bar{B}_s^0\rightarrow J/\psi K^{*0}$, the rates
obtained for these decays, relative to one of them to eliminate the factor $V'_p$, were in very good agreement with experiment~\cite{pdg}. 
The same conclusions were reached in the study of the $\bar{B}^0\rightarrow D^0\rho$ and $\bar{B}^0_s\rightarrow D^0K^{*0}$ in 
\cite{liangxie} and in the study of the semileptonic $D^+_s$, $D^+$, and $D^0$ into $\rho$, $\omega$, $K^*$, and $\phi$ done in \cite{Sekihara:2015iha}.
 In view of this, we proceed to convert the vector mesons
$\rho^0$, $\omega$, $\phi$ into a photon in order to have $J/\psi\gamma$ in the final state. 

For this we need the Lagrangian for this conversion, that is obtained from the local hidden gauge general Lagrangian~\cite{hidden1,hidden2,hidden4} as \cite{nagahiro}
\begin{equation}\label{Lvmd}
\mathcal{L}_{V\gamma}=-M_V^2\frac{e}{g}A_\mu\langle V^\mu Q\rangle,
\end{equation}
where $e$ is the electron charge, $e^2/(4\pi)=\alpha=1/137$, $g$ is the universal coupling in the local hidden gauge Lagrangian, $g=\frac{M_V}{2f_\pi}$, with
$M_V$ the vector meson mass, which we take as $M_V=800$ MeV, and $f_{\pi}=$ 93 MeV the pion decay constant. In Eq.~(\ref{Lvmd}), $A_\mu$, $V^\mu$ are the photon and vector meson fields and $Q=1/3(2,-1,-1)$ is the charge
matrix of the $u$, $d$, and $s$ quarks. 

\begin{figure}
  \centering
  \includegraphics[width=0.4\textwidth]{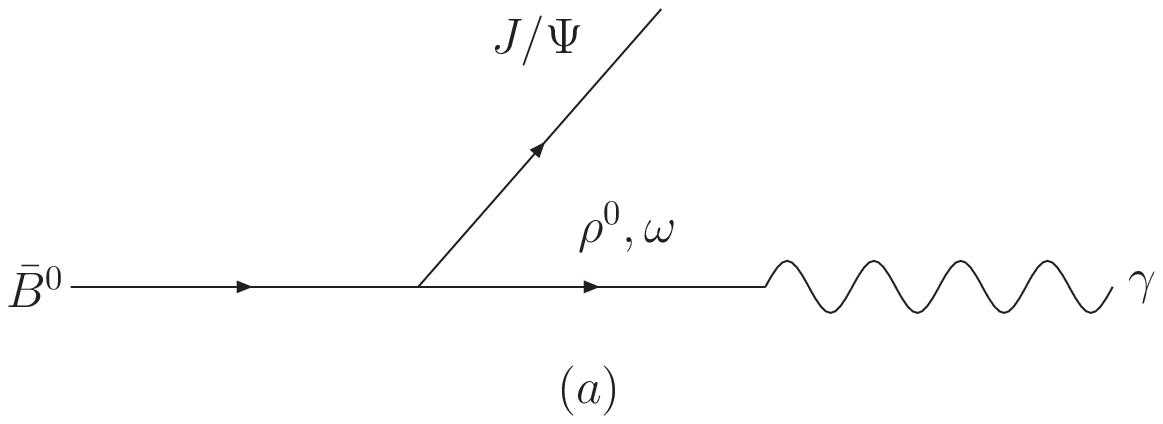}\;\;\;
 \includegraphics[width=0.4\textwidth]{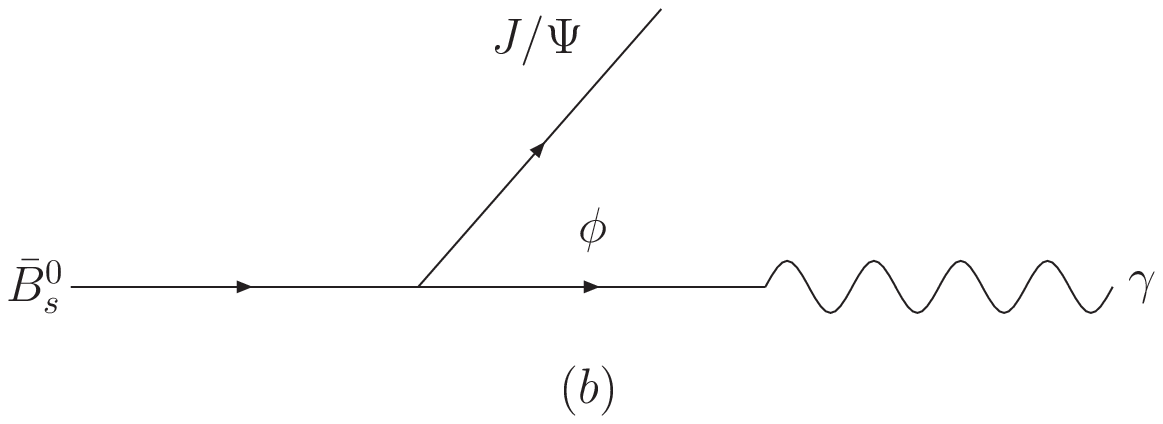}
  \caption{Diagrams for (a) $\bar{B}^0\rightarrow J/\psi\gamma$  and (b) $\bar{B}^0_s\rightarrow J/\psi\gamma$ (b) using vector meson dominance.}\label{VMD}
\end{figure}

The diagrams for the $\gamma$ production are now shown in Fig.~\ref{VMD}. The vector meson field in Eq.~(\ref{Lvmd}) is an SU(3) matrix and the symbol $\langle\rangle$ stands 
for the trace of $V^\mu Q$. The field $V^\mu$ is given by
\begin{equation}
V^\mu=\left(\begin{array}{ccc}
\frac{\rho^0}{\sqrt{2}}+\frac{\omega}{\sqrt{2}} &\rho^+ & K^{*+}\\
\rho^- & -\frac{\rho^0}{\sqrt{2}}+\frac{\omega}{\sqrt{2}}& K^{*0}\\
K^{*-}& \bar{K}^{*0} & \phi
\end{array}\right)
\end{equation}
and then the $V\gamma$ Lagrangian can be written more simply as
\begin{equation}\label{VMDVgamma}
\mathcal{L}_{V\gamma}=-M_V^2\frac{e}{g} A_\mu\tilde{V}^\mu C_{\gamma V},
\end{equation}
where now $\tilde{V}^\mu$ stands for the $\rho^0$, $\omega$, $\phi$ fields and
\begin{equation}
C_{\gamma V}=\left\{\begin{array}{ll}
\frac{1}{\sqrt{2}} & \mbox{for $\rho^0$}\\
\frac{1}{3}\frac{1}{\sqrt{2}}&\mbox{for $\omega$}\\
-\frac{1}{3} &\mbox{for $\phi$}\end{array}
\right. .
\end{equation}

The matrix elements for the diagrams of Fig.~\ref{VMD} are given by
\begin{equation}
-it_{\bar{B}^0\rightarrow J/\psi V(V\rightarrow \gamma)}=-i t_{\bar{B}^0\rightarrow J/\psi V} \epsilon_\mu(V) \epsilon_\nu(V)\epsilon^\nu(\gamma)\frac{i}{q^2-M_V^2}(-i)M_V^2\frac{e}{g}C_{\gamma V}.
\end{equation}
A bit of algebra, summing over the $V$ polarization, yields
\begin{eqnarray}
t_{\bar{B}^0\rightarrow J/\psi V(V\rightarrow \gamma)}&=&-t_{\bar{B}^0\rightarrow J/\psi V} (-g_{\mu\nu}+\frac{p_V^\mu p_V^\nu}{M_V^2})\epsilon^\nu(\gamma)\frac{e}{g}C_{\gamma V}\nonumber\\
&=&t_{\bar{B}^0\rightarrow J/\psi V} \epsilon_\mu(\gamma)\frac{e}{g}C_{\gamma V},\label{tBgamma}
\end{eqnarray}
where $p_V^\mu$ stands for the vector (equal to the photon) momentum, and for the moment we do not put the structure of the $\bar{B}\rightarrow J/\psi V$ in terms of the vector polarization. We simply show that the polarization of $V$ gets converted
into the one of the photons with some factors. Omitting the  polarization of the photon in Eq.~(\ref{tBgamma}), as we have done in Eqs.~(\ref{BJV1})--(\ref{BJV3}), we can write

\begin{equation}
t_{\bar{B}^0\rightarrow J/\psi\gamma}=V'_p V_{cd}\frac{e}{g}\left(-\frac{1}{\sqrt{2}}\frac{1}{\sqrt{2}}+\frac{1}{\sqrt{2}}\frac{1}{3}\frac{1}{\sqrt{2}}\right)
=V'_p V_{cd}\frac{e}{g}\left(-\frac{1}{3}\right),
\end{equation}
\begin{eqnarray}
t_{\bar{B}^0_s\rightarrow J/\psi\gamma}=V'_p V_{cs}\frac{e}{g}\left(-\frac{1}{3}\right).
\end{eqnarray}

The decay widths for $\bar{B}^0(\bar{B}^0_s)\rightarrow J/\psi V$ and $\bar{B}^0(\bar{B}^0_s)\rightarrow J/\psi \gamma$ are given by
\begin{eqnarray}
\Gamma_{\bar{B}^0_s\rightarrow J/\psi\phi}&=&\frac{1}{8\pi}\frac{1}{m_{\bar{B}_s^0}^2} (V'_p)^2 V_{cs}^2 p_\phi,\\
\Gamma_{\bar{B}^0_s\rightarrow J/\psi\gamma}&=&\frac{1}{8\pi}\frac{1}{m_{\bar{B}_s^0}^2} |t_{\bar{B}^0_s\rightarrow J/\psi \gamma}|^2p_\gamma,
\end{eqnarray}
and similar ones for $\bar{B}^0$ decays, where $p_\phi$, $p_\gamma$ are the absolute value of the $\phi$, $\gamma$   momentum in the $\bar{B}^0_s$ rest frame.

Since we do not explicitly evaluate $V'_p$, we perform ratios of widths with respect to $\Gamma_{\bar{B}^0_s\rightarrow J/\psi\phi}$, and we take $\Gamma_{\bar{B}_s^0\rightarrow J/\psi\phi}$ from
experiment~\cite{pdg}, i.e.,
\begin{equation}\label{exbra}
\Gamma_{\bar{B}^0_s\rightarrow J/\psi\phi}=(1.00^{+0.32}_{-0.18})\times10^{-3}\times\Gamma_{\bar{B}^0_s}.
\end{equation}

Hence, the ratios we are interested in are
\begin{equation}
\frac{\Gamma_{\bar{B}^0\rightarrow J/\psi\gamma}}{\Gamma_{\bar{B}^0_s\rightarrow J/\psi\phi}}=\frac{V_{cd}^2}{V_{cs}^2}\left(\frac{e}{g}\right)^2\frac{1}{9}\frac{p_\gamma}{p_\phi}\label{rate1}
\left(\frac{m_{\bar{B}^0_s}}{m_{\bar{B}^0}}\right)^2,
\end{equation}
\begin{equation}
\frac{\Gamma_{\bar{B}^0_s\rightarrow J/\psi\gamma}}{\Gamma_{\bar{B}^0_s\rightarrow J/\psi\phi}}=\frac{1}{9}\left(\frac{e}{g}\right)^2\frac{p_\gamma}{p_\phi}.\label{rate2}
\end{equation}
Taking into account the CKM matrix elements, $V_{cd}=-\sin\theta_c=-0.22534$, $V_{cs}=\cos\theta_c=0.97427$, we obtain
\begin{equation}
\frac{\Gamma_{\bar{B}^0\rightarrow J/\psi\gamma}}{\Gamma_{\bar{B}^0_s}}=\left(3.32^{+1.06}_{-0.60}\right)\times10^{-8}\label{branchB0},
\end{equation}
\begin{equation}
\frac{\Gamma_{\bar{B}^0_s\rightarrow J/\psi\gamma}}{\Gamma_{\bar{B}_s^0}}=\left(6.21^{+2.00}_{-1.12}\right)\times10^{-7}\label{branchBS0},
\end{equation}
where the errors are the same relative errors of Eq.~(\ref{exbra}). For practical purposes, we can take $\Gamma_{\bar{B}^0_s}=\Gamma_{\bar{B}^0}$ (which differ only by a few percent~\cite{pdg}) and, thus,
Eqs.~(\ref{branchB0})--(\ref{branchBS0}) give branching ratios. 

It is interesting to compare the results of Eqs.~(\ref{branchB0})--(\ref{branchBS0}) with the present experimental bounds for these branching ratios. The LHCb's recent work ~\cite{lhcb} quotes 
at 90\% confidence level
\begin{equation}
\frac{\Gamma_{\bar{B}^0\rightarrow J/\psi\gamma}}{\Gamma_{\bar{B}^0}}< 1.5\times10^{-6}\label{lhcb1},
\end{equation}
\begin{equation}
\frac{\Gamma_{\bar{B}^0_s\rightarrow J/\psi\gamma}}{\Gamma_{\bar{B}^0_s}}< 7.3\times10^{-6}\label{lhcb2}.
\end{equation}
As we can see, the results that we obtain from Eqs.~(\ref{branchB0})--(\ref{branchBS0}) are consistent with the experimental bounds of Eqs.~(\ref{lhcb1})--(\ref{lhcb2}). It is interesting to note
that the rate we obtain for $\Gamma_{\bar{B}^0\rightarrow J/\psi\gamma}$ is much smaller than the present bound, but the rate that we
get for $\Gamma_{\bar{B}^0_s\rightarrow J/\psi\gamma}$ is only one order of magnitude smaller than the present bound.

\begin{figure}[htpb]
  \centering
  \includegraphics[width=0.8\linewidth]{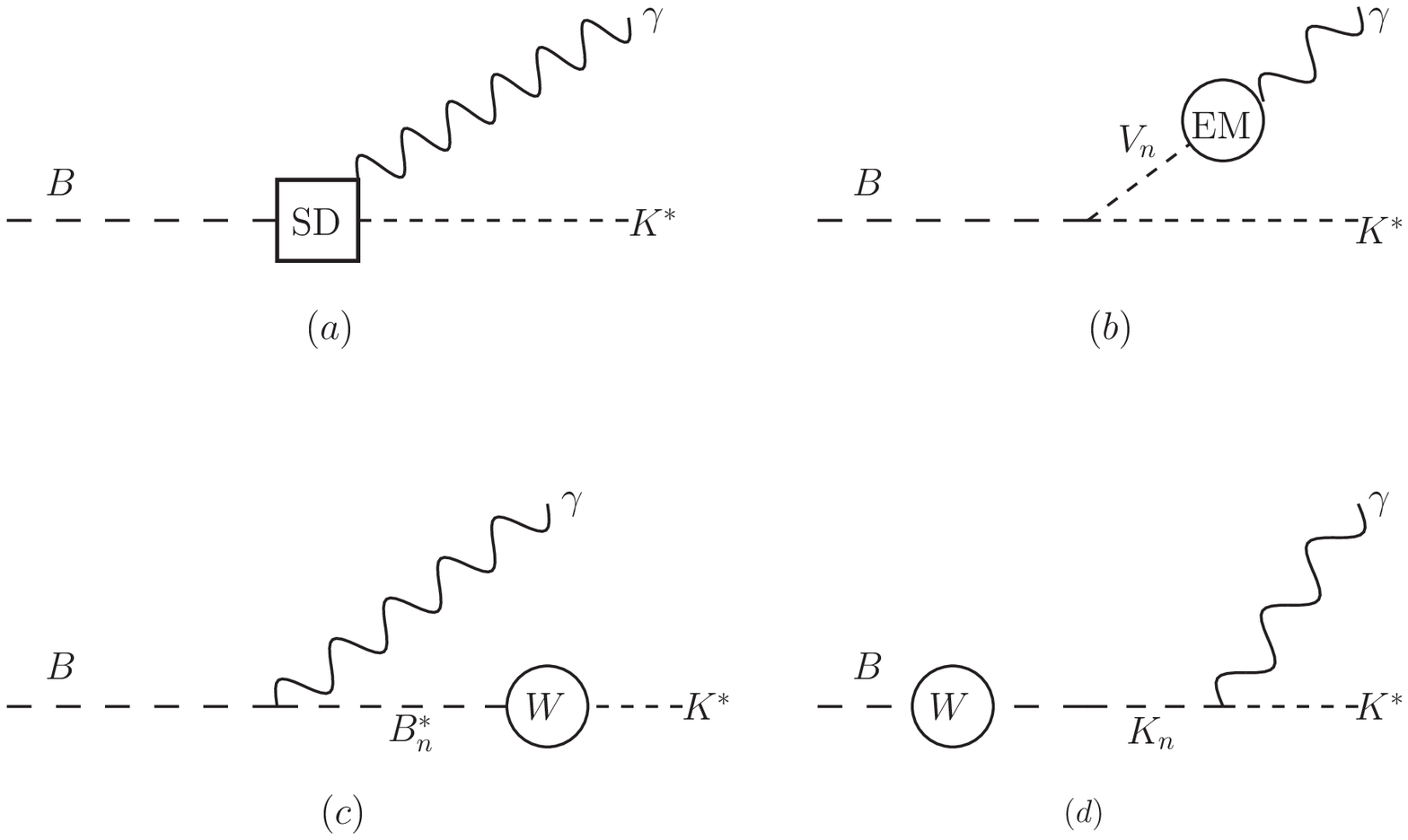}
  \caption{Diagrams for $B\rightarrow \gamma K^*$: (a) short range processes; (b)-(d) long range processes ($n$ indicates possible intermediate states).}
  \label{Fig:new}
\end{figure}

\begin{figure}[htpb]
  \centering
  \includegraphics[width=0.8\linewidth]{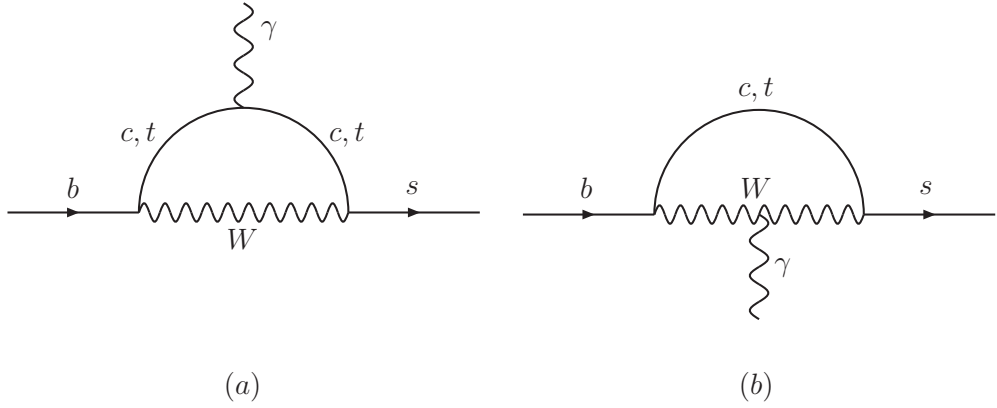}\;
  \caption{Basic diagrams involving the $b\rightarrow s\gamma$ transition responsible for the short range part in $B\rightarrow \gamma K^*$.}
  \label{Fig:short}
\end{figure}

\begin{figure}[htpb]
  \centering
  \includegraphics[width=0.8\linewidth]{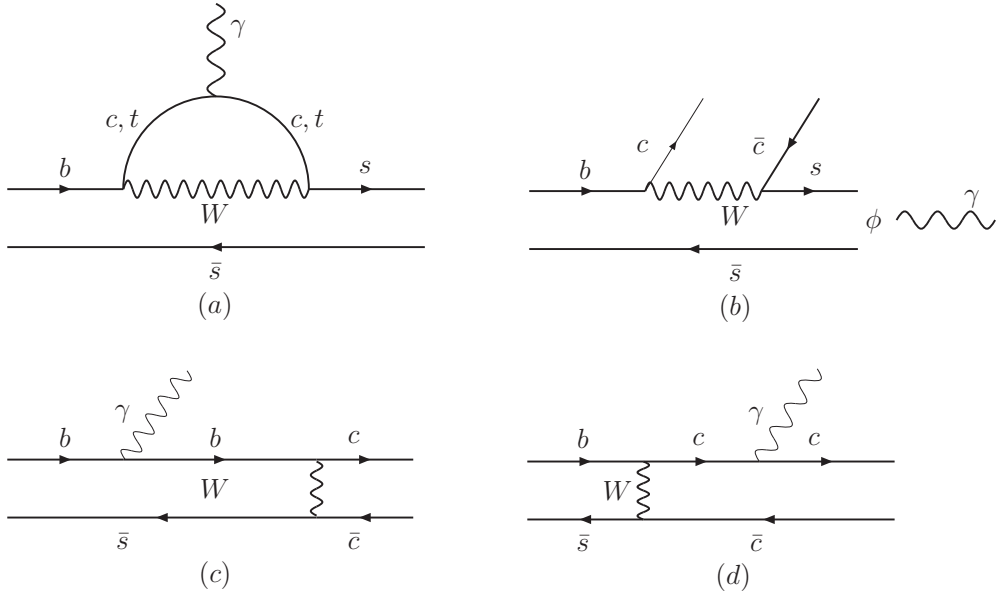}\;
  \caption{Diagrams of Fig.~\ref{Fig:new} made explicit for $\bar{B}^0_s\rightarrow J/\psi\gamma$.}
  \label{Fig:meca}
\end{figure}

At this point, and before we go into a discussion of the spin structure of the amplitudes in the next section, we would like to 
situate the work into a more general context. The mechanism that we use for the decay classifies into what is denoted as long range processes in
\cite{golo,cheng,golo2,dono,burdman}. In these works, the processes $B\rightarrow\gamma K^*$, $B\rightarrow\gamma\rho$ have been studied and
the mechanisms are separated into a short range part and a long range part. The long range part is evaluated using the concept of vector meson dominance, much
as it has been done here. Schematically, the mechanisms are depicted in Fig.~\ref{Fig:new} for the $B\rightarrow \gamma K^*$ decay.  In \cite{golo,cheng,golo2} it was found that the short range diagram (a) dominated the amplitude. The explicit mechanism responsible for the large contribution of diagram (a) is depicted in
Fig.~\ref{Fig:short}~\cite{bertolini,grinstein,burdman}.  The penguin diagrams of Fig.~\ref{Fig:short} are dominated by the two-quark 
intermediate states~\cite{burdman} and they lead to a branching fraction $\mathcal{B}(B^0\rightarrow \gamma K^{*0})=(4.33\pm0.15)\times10^{-5}$~\cite{pdg}. The rate is about a factor 30 larger than the boundary for $\bar{B}^0\rightarrow J/\psi\gamma$ quoted in Eq.~(\ref{lhcb1}), indicating that the equivalent short range mechanism
might be absent in the $\bar{B}^0\rightarrow J/\psi\gamma$ reaction. This would not be an exception since in \cite{burdman} it was found that
the short range terms are much smaller than those of the long range in the radiative decay of charm mesons. In order to shed some light on this issue, we plot in Fig.~\ref{Fig:meca} the four mechanisms of Fig.~\ref{Fig:new}  for $\bar{B}_s\rightarrow J/\psi\gamma$ considering the explicit form of Fig.~\ref{Fig:short} for the short range mechanism.  We can see that  diagram (a), which includes the $b\rightarrow s\gamma$ transition that was found to have a large value in \cite{golo,cheng,golo2,dono,burdman} for
the $B\rightarrow\gamma K^*$ transition, does not lead to $J/\psi$ in the final state. It would instead contribute to $\bar{B}_s\rightarrow \gamma\phi$ and actually we can see that the branching fraction for this mode is indeed large, $\mathcal{B}(B^0_s\rightarrow\gamma\phi)=(3.52\pm0.34)\times10^{-5}$.  On the other hand,
diagrams (b), (c), and (d), described as long range in \cite{golo}, all  can lead to $J/\psi$ in the final state through the combination of $c\bar{c}$. The diagram that we have calculated corresponds to diagram (b). With this perspective we can justify the suppression of the mechanisms of (c) and (d) with respect to (b).
Indeed, diagram (b) has the weak process in just one quark of the original $\bar{B}_s$,  while (c) and (d) involve two quarks. These processes, including two body matrix elements, are usually penalized with respect to those including only one body (see discussions in Sec. 4 of \cite{dai}). On the other hand, in diagram (c), one has an intermediate $b\bar{s}$ state which is off shell by the energy carried out by the photon (about 1.7 GeV), and in diagram (d) the $c\bar{c}$ intermediate state is off shell by about the same amount. The double penalization should make these two mechanisms small compared to diagram (b), which would be the dominant term for this reaction. 

\section{Polarization structure of the vertices and comparison with other works}
So far, we have not paid attention to the structure of the $B\rightarrow J/\psi\gamma$ vertex. In fact, we can have
two possible structures, one that conserves parity (PC) and another one that violates parity (PV), both of which are allowed in the weak decay. The structures
are
\begin{equation}\label{vpc}
V_{PC}=\epsilon_{\mu\nu\alpha\beta}\epsilon^\mu(J/\psi)q^\nu(J/\psi)\epsilon^{'\alpha}(V)q^{'\beta}(V)
\end{equation}
where $q$, $q'$ are the momenta of the $J/\psi$ and $V$, respectively. For the case of photon production, $\epsilon'$ and $q'$ will then stand for the photon. The other structure is given by
\begin{equation}\label{vpv}
V_{PV}=\epsilon^\mu(J/\psi)\epsilon^{'\nu}(V)(g_{\mu\nu} q\cdot q'-q'_\mu q_\nu)
\end{equation}
and again $\epsilon'$, $q'$ would be the polarization and momenta of the photon for the case of photon production. Note that in the case of photon production the two structures are gauge invariant. These two structures are explicitly used in
the theoretical works~\cite{golo2,luwang,lilu,anas}.

For the case of $V\gamma$ in the final state, the two structures guarantee gauge invariance, but for the case of $VV$ such restriction is not necessary in principle.
This issue was widely discussed in \cite{golo2} since by starting with a more general amplitude for $VV$ and implementing the vector meson dominance (VMD) $V\gamma$ conversion
Lagrangian of Eq.~(\ref{VMDVgamma}), the resulting amplitude might not be gauge invariant. Some prescription is given in \cite{golo2}, eliminating the longitudinal-longitudinal $VV$ helicities in the $VV$ process and then applying the VMD conversion. While this can be a reasonable approach, we would like to recall that a systematic study
of the $VV$ and $V\gamma$ processes using the local hidden gauge approach~\cite{hidden1,hidden2,hidden4} to deal with vector mesons and their interaction produces gauge invariant amplitudes after the $V\gamma$ conversion via Eq.~(\ref{VMDVgamma}). This comes after subtle cancellations due to a contact term and vector exchange interactions. This has been shown
explicitly in the radiative decay of axial vector mesons in \cite{nagahiro} and in the $\gamma\gamma$ decay of the $f_0(1370)$ and $f_2(1270)$ resonances \cite{junko}. In view of this, and to guarantee the forms of Eqs.~(\ref{vpc}) and (\ref{vpv}) for the case of $B\rightarrow J/\psi \gamma$ decay, we assume the same structure
for the $VV$ decay, which leads to Eqs.~(\ref{vpc})--(\ref{vpv}) upon the VMD transition of Eq.~(\ref{VMDVgamma}).  In \cite{dono} a final state interaction of the $\rho\rho$ in the $B\rightarrow \rho\rho\rightarrow\rho\gamma$ process is taken into account. We do not do that explicitly since this would be accounted for by our $B\rightarrow VV$ amplitude of 
Eqs.~(\ref{vpc})--(\ref{vpv}). Explicit evaluations of this interaction are done in \cite{junko} following the $VV$ interaction of the local hidden gauge approach in \cite{raquel,gengvec}.
Nonetheless, in our case with $J/\psi$-light vector meson interaction, this interaction is weak and proceeds through coupled channels, since the tree level $J/\psi V$ interaction is zero because we cannot exchange a $q\bar{q}$ pair from the $c\bar{c}$ pair to the light vectors.

In the evaluation of the rates of $\bar{B}^0(\bar{B}^0_s)\rightarrow J/\psi V$ in \cite{bayarliang}, the explicit structure of the vertices was irrelevant, as far as one takes the $V$ vector masses to be
equal, which is a good approximation. However, here, the structures can give some different weights depending on whether one has a vector meson or a photon in the final state.  Hence we  evaluate the weights
of these structures for the particular case that we have. We find after summing over polarization of the vector mesons or the photon (we get the same structure in both cases)
\begin{equation}\label{sum1}
\sum\limits_\lambda\sum\limits_{\lambda'}|V_{PC}|^2=2((q\cdot q')^2-q^2 q^{'2})
\end{equation}

\begin{equation}\label{sum2}
\sum\limits_\lambda\sum\limits_{\lambda'}|V_{PV}|^2=2(q\cdot q')^2 + q^2 q^{'2}
\end{equation}
where $q^2=M_{J/\psi}^2$, $q'^2=M_V^2$ or 0 (for $V$ or $\gamma$ production),
and
\begin{equation}
q\cdot q'=\frac{1}{2}\left(M_{\bar{B}^0(\bar{B}^0_s)}^2-M^2_{J/\psi}-M_{V(\gamma)}^2\right)
\end{equation}

The fact that Eqs.~(\ref{vpc})--(\ref{vpv}) are gauge invariant guarantees that only the transverse polarizations of the photon contribute.
This can be easily shown by explicitly taking the sum over the transverse photon polarizations
\begin{equation}\label{eq:photonpol}
\sum\limits_\lambda\epsilon_i(\gamma)\epsilon_j(\gamma)=\delta_{ij}-\frac{q'_i q'_j}{\vec{q'}^2}
\end{equation}
instead of the covariant one $\sum_\lambda \epsilon_\mu\epsilon_\nu\approx-g_{\mu\nu}$  valid for gauge invariant amplitudes. In both cases, one reproduces
the results of Eqs.~(\ref{sum1})--(\ref{sum2}).

So far, in the results we have shown in Eqs.~(\ref{branchB0})--(\ref{branchBS0}), the structures Eqs.~(\ref{sum1})--(\ref{sum2}) are not taken into account. In order to evaluate the ratios of Eqs.~(\ref{rate1})--(\ref{rate2}) taking into account
these vector structures, we would have to  multiply these ratios by the following ratio,  
\begin{equation}\label{Rfactor}
R=\frac{\sum\limits_\lambda\sum\limits_{\lambda'}|V_{PC(PV)}|^2\;\mbox{for photon}}
{\sum\limits_\lambda\sum\limits_{\lambda'}|V_{PC(PV)}|^2\;\mbox{for $\phi$}},
\end{equation}
which is shown in Table I.

\begin{table}
\setlength{\tabcolsep}{0.5cm} 
\centering
\caption{Values of the $R$ correction factor of Eq.~(\ref{Rfactor}).}\label{Table:origin}
\begin{tabular}{lll}
  \hline\hline
  R & $V_{PC}$ & ~~$V_{PV}$ \\\hline
  $\Gamma_{\bar{B}^0\rightarrow J/\psi\gamma}$ & 1.15 &           ~~ 0.95 \\
  $\Gamma_{\bar{B}^0_s\rightarrow J/\psi\gamma }$ & 1.27 & ~~1.05\\
  \hline\hline
\end{tabular}
\end{table}

Taking into account these correction factors as a source of systematic uncertainties, together with the statistical ones of Eq.~(\ref{exbra}), we obtain
\begin{equation}
\mathrm{Br}[\bar{B}^0\rightarrow J/\psi\gamma]=\left(3.50\pm0.34^{+1.12}_{-0.63}\right)\times10^{-8},\label{final1}
\end{equation}
\begin{equation}
\mathrm{Br}[\bar{B}^0_s\rightarrow J/\psi\gamma]=\left(7.20\pm0.68^{+2.31}_{-1.30}\right)\times10^{-7}.\label{final2}
\end{equation}

It is interesting to compare these results with other theoretical calculations~\cite{luwang,lilu,anas}. In \cite{luwang} the authors present two calculations: one of them uses the naive factorization and the other considers nonfactorizable contributions. In \cite{lilu}, the authors use perturbative QCD based on $K_T$ factorization, where the $\mathrm{Br}(\bar{B}^0\rightarrow J/\psi\gamma)$ rate is evaluated
explicitly and then the $\mathrm{Br}[\bar{B}^0_s\rightarrow J/\psi\gamma]$ is obtained using SU(3) arguments. In \cite{anas} the factorization approximation
is used, taking into account the photon emission not only from the $B$-meson loop but also from the vector-meson loop. All these results, together with ours, are shown in Table II.  One can easily notice the large variation among the results, which can differ by two orders of magnitude.  The results that we obtain
for the two decay rates are in the middle of the other theoretical results. It should be noted that in our approach, the relatively small
$\mathrm{Br}[\bar{B}^0\rightarrow J/\psi\gamma]$ compared with $\mathrm{Br}[\bar{B}^0_s\rightarrow J/\psi\gamma]$  can be traced back
to the ratio $|V_{cd}/V_{cs}|^2\approx1/20$, like in \cite{lilu}. 
\begin{table}
\centering
\setlength{\tabcolsep}{0.4cm} 
\caption{Values of different theoretical evaluations.}\label{Table:origin}
\begin{tabular}{lcc}
  \hline\hline
   Models & $\mathrm{Br}[\bar{B}^0\rightarrow J/\psi\gamma]$ & $\mathrm{Br}[\bar{B}^0_s\rightarrow J/\psi\gamma]$ \\\hline
  Naive factorization~\cite{luwang} & $5.40\times10^{-8}$ & $1.40\times 10^{-6}$\\
  QCD factorization~\cite{luwang} & $ 2.44\times10^{-9}$ & $5.80\times10^{-8}$\\
  Perturbative QCD $K_T$ factorization~\cite{lilu} & $4.5\times 10^{-7}$ & $5.0\times 10^{-6}$\\
  Factorization approach~\cite{anas} & $7.54\times10^{-9}$ & $1.43\times 10^{-7}$\\
  This work & $\left(3.50\pm0.34^{+1.12}_{-0.63}\right)\times 10^{-8}$ & $\left(7.20\pm0.68^{+2.31}_{-1.30}\right)\times 10^{-7}$\\
  \hline\hline
\end{tabular}
\end{table}

One should note that the approaches seem totally different, but they are not so. The elaborate calculations done in \cite{luwang,lilu,anas} would go in our approach in the evaluation of $\bar{B}^0(\bar{B}^0_s)\rightarrow J/\psi V$ which we do not do explicitly.  Instead we use the experimental value of $\bar{B}^0_s\rightarrow J/\psi \phi$. Then, with the help of Ref.~\cite{bayarliang}, where one relates theoretically the different $\bar{B}^0(\bar{B}^0_s)\rightarrow J/\psi V$ decays, and the VMD hypothesis, we can evaluate finally the rates of Eqs.~(\ref{final1})--(\ref{final2}). One should note that the form factors used in \cite{luwang,lilu,anas} also rely on some other 
observables for their determination. In this sense, it is not so much the physics, but the strategy to get the rates, which changes from our approach to the other ones. The fact that we obtained very good rates for the $\bar{B}^0(\bar{B}^0_s)\rightarrow J\psi V$ decays in \cite{bayarliang} and the reliability of the VMD in the range of energies studied here should made our predictions rather solid. Indeed, explicit application of the local
hidden gauge approaches and vector meson dominance gives good rates for  $f_2(1270)\rightarrow\gamma\gamma$ and $f_0(1370)\rightarrow\gamma\gamma$~\cite{junko}, the two-photon and one-photon-one-vector decay widths of $f_0(1370)$, $f_2(1270)$, $f_0(1710)$, $f'_2(1525)$ and $K^*_2(1430)$~\cite{branzgeng}, and others~\cite{talk}.

It is interesting to note that the branching ratio that we get for $\bar{B}^0_s\rightarrow J/\psi \gamma$ is just one order of magnitude smaller than the experimental bound. With increasing statistics in present
facilities, this should serve as an incentive for extra experimental efforts in this reaction to determine an absolute rate.

\section{loop corrections}
\begin{figure}[htpb]
  \centering
  \includegraphics[width=0.4\linewidth]{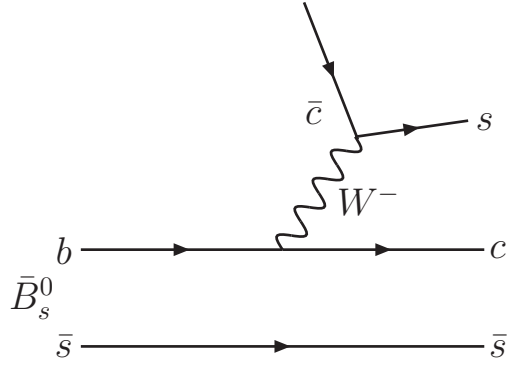}\;
  \caption{Color favored $\bar{B}^0_s\rightarrow D^+_s D^-_s$ mechanism.}
  \label{Fig:loop1}
\end{figure}

\begin{figure}[htpb]
  \centering
  \includegraphics[width=1.0\linewidth]{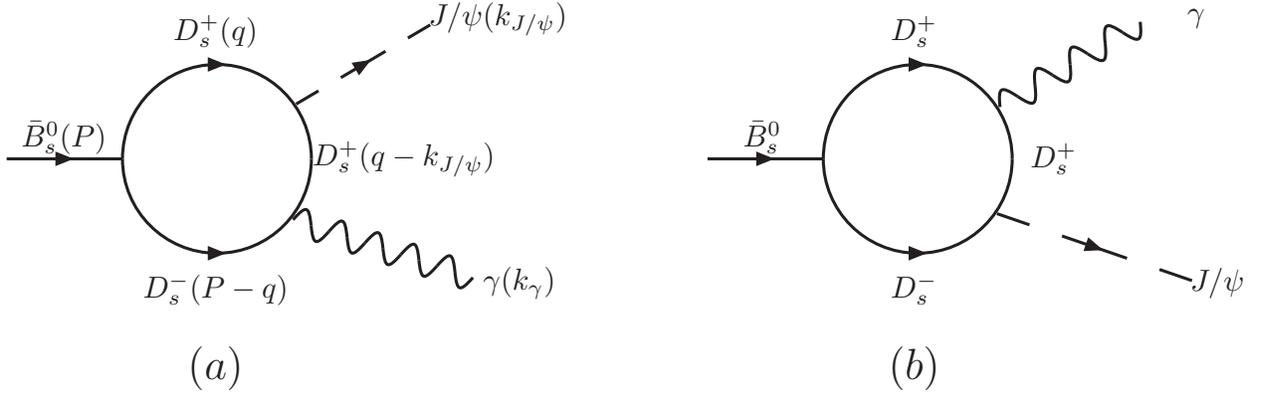}\;
  \caption{Loop diagrams for $\bar{B}^0_s\rightarrow J/\psi\gamma$ through $D_s^+D_s^-$ intermediate production. In diagram (a), the four momenta
  of the particles are given in the parentheses.}
  \label{Fig:loop2}
\end{figure}
The mechanism of Fig. 1(b) for $\bar{B}^0_s\rightarrow J/\psi (s\bar{s})$ is color suppressed compared with the mechanism  for 
$\bar{B}^0_s\rightarrow D_s^+ D_s^-$ depicted in Fig.~\ref{Fig:loop1}, which is color favored.  In view of this, one may wonder why the loop correction $\bar{B}^0_s\rightarrow D^+_s D^-_s\rightarrow J/\psi\gamma$ could not compete with the tree level process studied so far. To answer this question we evaluate the contribution of the
loop of Fig.~\ref{Fig:loop2}.

The evaluation requires the use of the Lagrangians
\begin{equation}
\mathcal{L}_{VPP}=-ig\langle [\phi,\partial_\mu\phi]V^\mu\rangle,
\end{equation}
where $g=\frac{M_V}{2 f_\pi}$ with $M_V\approx800$ MeV and $f_\pi=93$ MeV. The matrices $\phi$ and $V^\mu$ are extended to SU(4) to accommodate 
the $D_s$ and $J/\psi$ mesons and are given in \cite{danizou}. As a consequence we find 
\begin{equation}
-i t_{J/\psi, D_s^+D_s^-}\equiv i g\epsilon^\mu_{J/\psi}(2 q- k_{J/\psi})_\mu. 
\end{equation}
The coupling of the photon to the pseudoscalar mesons is equally given by
\begin{equation}
-i t_{\gamma,D_s^+D_s^-}=i e\epsilon^\mu_\gamma (2q-k_{J/\psi}-P)_\mu,
\end{equation}
with $e^2/(4\pi)=\alpha\approx1/137$.

Since there is much phase space for the $\bar{B}^0_s\rightarrow D^+_s D^-_s$ decay, the loop function is dominated by the positive energy part of the
$D^+_s$,  $D^-_s$ propagators emerging from the $\bar{B}^0_s$ and the loop function is also dominated by its imaginary part~\cite{npa}. Then we can write 
in the rest frame of the $\bar{B}^0_s(\vec{P}=0)$
\begin{eqnarray}
t&=&i\int\frac{d^4 q}{(2\pi)^4} e g_B g \frac{1}{2\omega(\vec{q})}\frac{1}{q^0-\omega(\vec{q})+i\epsilon}\nonumber\\
&\times&\frac{1}{2\omega(\vec{q})}
\frac{1}{P^0-q^0-\omega(\vec{q})+i\epsilon}\frac{1}{q^0-k_{J/\psi}^0-\omega'(\vec{q})+i\epsilon}\frac{1}{q^0-k^0_{J/\psi}+\omega'(\vec{q})-i\epsilon} \nonumber\\
&\times& \epsilon^\mu_{J/\psi}(2q-k_{J/\psi})_\mu \epsilon_\gamma^\nu (2q-k_{J/\psi}-P)_\nu F(\vec{q}-\vec{k}_{J/\psi})
\end{eqnarray}
with $g_B$ the coupling of $\bar{B}^0_s\rightarrow D_s^+ D_s^-$, $\omega(\vec{q})=\sqrt{\vec{q}\,^2+m_{D_s}^2}$, $\omega'(\vec{q})=\sqrt{(\vec{q}-\vec{k}_{J/\psi})^2+m_{D_s}^2}$, and $F(\vec{q}-\vec{k}_{J/\psi})$ a form factor to account for the off-shellness of the $J/\psi$ and $\gamma$ couplings with the
$(q-k_{J/\psi})$ $D_s$ meson off shell. We take an empirical coupling of the type
\begin{equation}
F(\vec{p})=\frac{\Lambda^2}{\Lambda^2+\vec{p}^2}
\end{equation}
and $\Lambda=1$ GeV or less. 

By performing the $q^0$ integration analytically we obtain 
\begin{eqnarray}
t&=&e g_B g \epsilon_i(J/\psi)\epsilon_j(\gamma)\int\frac{d^3 q}{(2\pi)^3}\frac{1}{2\omega(\vec{q})}\frac{1}{2\omega(\vec{q})}\frac{1}{2\omega'(\vec{q})}\nonumber\\
&\times&\frac{1}{P^0-\omega(\vec{q})-\omega(\vec{q})+i\epsilon}\frac{1}{P^0-k^0_{J/\psi}-\omega(\vec{q})-\omega'(\vec{q})+i\epsilon}\frac{1}{\omega(\vec{q})+\omega'(\vec{q})-k^0_{J/\psi}-i\epsilon}\nonumber\\
&\times&(2\omega(\vec{q})+2\omega'(\vec{q})-P^0)(2 q-k_{J/\psi})_i (2q-k_{J/\psi})_j,
\end{eqnarray}
where we keep explicitly the $\gamma$ polarization vector spatial and we also neglect the three-momentum of the $J/\psi$  versus its mass. The integral
gives a result of the type $a\delta_{ij}+b k_{\gamma,i} k_{\gamma,j}$ ($\vec{k}_\gamma=-\vec{k}_{J/\psi}$) but the 
second term vanishes with transverse photons. The second diagram of Fig.~\ref{Fig:loop2} gives the same contribution as the first one and, considering
explicitly that we have only transverse photons, we have
\begin{equation}
t=4 e g_B g \vec{\epsilon}(J/\psi)\vec{\epsilon}(\gamma) I
\end{equation}
and the sum over polarizations, taking Eq.~(\ref{eq:photonpol}) into account, gives
\begin{equation}
\bar{\sum}\sum |t|^2=32 e^2 g_B^2 g^2 |I|^2,
\end{equation}
with 
\begin{eqnarray}
I&=&\int\frac{d^3 q}{(2\pi)^3}\sin^2\theta\left(\frac{1}{2\omega(\vec{q})}\right)^2\frac{1}{2\omega'(\vec{q})}\frac{1}{P^0-2\omega(\vec{q})+i\epsilon}F(\vec{q}-\vec{k}_{J/\psi})\nonumber\\
&\times&\frac{1}{P^0-k^0_{J/\psi}-\omega(\vec{q})-\omega'(\vec{q})+i\epsilon}\frac{1}{\omega(\vec{q})+\omega'(\vec{q})-k^0_{J/\psi}-i\epsilon}(2\omega(\vec{q})+2\omega'(\vec{q})-P^0)
\end{eqnarray}
By taking the imaginary part of $I$ we find
\begin{eqnarray}\label{eq:imI}
i\mathrm{Im} I&=&-i\frac{1}{8\pi} q^3_\mathrm{on}\frac{1}{4\omega(q_\mathrm{on})}\int^1_{-1} d\cos\theta\sin^2\theta\frac{1}{2\omega'(\vec{q})}\nonumber\\
&\times&\frac{1}{P^0-k^0_{J/\psi}-\omega(\vec{q})-\omega'(\vec{q})+i\epsilon}\frac{1}{\omega(\vec{q})+\omega'(\vec{q})-k^0_{J/\psi}-i\epsilon}(2\omega(\vec{q})+2\omega'(\vec{q})-P^0)
\end{eqnarray}
with $|\vec{q}|=q_\mathrm{on}$, and $q_\mathrm{on}$ the $D_s$ on-shell momentum for $\bar{B}^0_s\rightarrow D_s^+ D_s^-$. 
The two denominators in Eq.~(\ref{eq:imI}) do not lead to poles in $\mathrm{Im} I$.  The coupling $g_B$ of $\bar{B}^0_s$ to $D_s^+ D_s^-$ is
taken from experiment~\cite{pdg} and we have
\begin{equation}
\Gamma_{\bar{B}^0_s\rightarrow D^+_s D^-_s}=\frac{p_{D_s}}{8\pi M_{\bar{B}^0_s}^2}g_B^2,
\end{equation}
from which
\begin{equation}
\frac{g_B^2}{\Gamma_B}=\frac{8\pi M_{\bar{B}^0_s}^2}{p_{D_s}}\frac{\Gamma_{\bar{B}^0_s\rightarrow D^+_s D^-_s}}{\Gamma_{\bar{B}^0_s}}
\end{equation}
and
\begin{equation}
\frac{\Gamma_{\bar{B}^0_s\rightarrow D^+_s D^-_s}}{\Gamma_{\bar{B}^0_s}}=4.4\times10^{-3}\quad\quad\mbox{\cite{pdg}}.
\end{equation}
Altogether we find now
\begin{equation}
\frac{\Gamma_{\bar{B}^0_s\rightarrow\gamma J/\psi}}{\Gamma_{\bar{B}^0_s}}=\frac{p_\gamma}{p_{D_s}}\frac{\Gamma_{\bar{B}^0_s\rightarrow D^+_s D^-_s}}{
\Gamma_{\bar{B}^0_s}}32 e^2 g^2|I|^2.
\end{equation}
By taking $\Lambda=1$ GeV, we find
\begin{equation}
\frac{\Gamma_{\bar{B}^0_s\rightarrow\gamma J/\psi}}{\Gamma_{\bar{B}^0_s}}\approx 4.8\times10^{-8},
\end{equation}
which is about a factor of 20 smaller than what we obtained from vector meson dominance in Eq.~(\ref{final2}). But taking $\Lambda=1.2$ GeV the branching ratio is
$7.98\times 10^{-8}$, still one order of magnitude smaller than what was found before. 

Certainly, one can think of similar loops with $D_s$, $D_s^*$ intermediate states, but the exercise done indicates that these loops should be reasonably 
smaller than what has been calculated before.

There is more to it: we can look at the diagrams of Fig.~\ref{Fig:loop2} and replace a $\gamma$ by a $\phi$ (or $J/\psi$) meson. In the vector meson dominance picture the $\gamma$ production amplitude  is obtained from an amplitude producing   $\phi$ and $J/\psi$ followed by conversion of $\phi$ and $J/\psi$ into a photon.
 If we ignore the $J/\psi$ contribution, the $\phi$ contribution
alone is already accounted for in our formalism, since we take the $\bar{B}^0_s\rightarrow J/\psi\phi$ process from experiment and convert the
$\phi$ into a $\gamma$. The empirical process also accounts for this loop contribution. Hence, what one is missing is only the fraction of the loop diagram that has the $\gamma$ formed from $J/\psi$. Their contributions have strength $\frac{1}{3}e$, $\frac{2}{3}e$ for $\phi$, $J/\psi$, summing to the $e$ charge,
and hence what is missing is still smaller than the loop function that we have calculated.

There is another empirical proof that these loop corrections are small. Indeed, as we have commented, replacing the $\gamma$ in Fig.~\ref{Fig:loop2} by a $\phi$ gives a contribution 
to $\phi$ production coming from loops. The same diagram does not contribute to $\rho$ and $\omega$ production, since neither of them couples to $D_s$. This means that the loop contributions are very selective to the vector mesons produced. If the loop corrections to $J/\psi V$ ($V=\rho,\omega,\phi$) were important, then one would not obtain good results for these processes omitting the loops. Yet, the works done in \cite{bayarliang,liangxie,Sekihara:2015iha} on $B$ and $D$ decays, taking only the tree level diagram 
of Fig. ~1 and considering the vector mesons as coming solely from the final $q\bar{q}$ pair, indicate that this picture is rather accurate for the ratios of branching ratios, in agreement with experiment within errors.

\section{Summary}
Radiative $B$ decays  are potentially sensitive to both the standard model physics  and beyond the standard model physics. Recent studies based on a novel nonperturbative mechanism, which includes a primary 
quark level transition, and the following hadronization and final state interactions of the produced hadrons, are  capable of explaining very successfully  a large variety of  experimental data with a minimum amount of input.
In the present work, we have extended such an approach and  utilized the vector meson dominance hypothesis to predict the branching ratios of the radiative $B$ decays. The resulting parameter-free predictions not only are
consistent with the present experimental upper bounds but also show a characteristic pattern that can be verified experimentally. Our results show that although the $\bar{B}^0 \rightarrow J/\psi\gamma$ decay rate
is too small to be detected in the near future, the $\bar{B}^0_s\rightarrow J/\psi\gamma$ is much closer to the capacity of the LHCb detector. These results should serve to encourage our experimental colleagues to
continue their efforts to obtain an absolute rate for this decay process. It should be stressed that unlike earlier theoretical studies based on QCD factorization or perturbative QCD, the approach developed in the present work
 relies on experimental information mostly and, as a result, should be free of uncertainties inherent in earlier studies.

\section{Acknowledgements}
L.S.G thanks the nuclear theory group of Valencia University for its hospitality during his visit. This work is partly supported by the National Natural Science Foundation of China under Grants No. 11375024 and No. 11522539, the Spanish Ministerio de Economia y Competitividad and European FEDER funds under Contract No.
FIS2011-28853-C02-01 and No. FIS2011-28853-C02-02, and the Generalitat Valenciana in the program Prometeo II-2014/068.

\end{document}